# Proximitization: Opportunities for Manipulating Correlations in Hybrid Organic / 2D Materials

Joohyung Park,[1] Ayan N. Batyrkhanov,[1] J.R. Schaibley[2] and Oliver L.A. Monti[1,2]

[1] Department of Chemistry and Biochemistry, The University of Arizona, 1306 E. University Blvd., Tucson, AZ, USA

[2] Department of Physics, The University of Arizona, 1118 E. Fourth Street, Tucson, AZ, USA

Email: monti@arizona.edu

Tel: +1 520 626 1177

# Abstract

Van der Waals layered and 2D materials constitute an extraordinary playground for condensed matter physics, since the strong confinement of wavefunctions to two dimensions supports a diverse set of correlated phenomena. By creating carefully designed heterostructures, these can be readily manipulated. In this Perspective, we advance the viewpoint that heterostructures from from these materials with thin layers of organic molecules offer an opportunity for creating and manipulating the correlated degrees of freedom in unprecedented ways. We briefly survey what has been accomplished thus far, including proposed mechanisms, before concentrating on unique opportunities offered by the vast selection of available organic molecules. We further introduce the notion of "proximitization" in combination with symmetry breaking as a fertile and potentially unifying conceptual vantage point from which to consider opportunities for tailoring correlations in van der Waals layered materials.

# I.    Introduction

The discovery of graphene[1] by Novoselov and Geim in 2004 and the subsequent extension to other atomically thin crystalline materials that can be obtained from van der Waals layered bulk materials marks the dawn of a new era in condensed matter physics. The ease with which this new class of materials can be prepared, e.g., by exfoliation to the 2D limit or by chemical vapor deposition, has enabled assembly of complex layered structures with atomically precise control. Remarkably, most phases and phenomena associated with strong electronic correlations (i.e. where a description beyond non-interacting single particles is needed) are represented among these materials: van der Waals materials span the range from metals to insulators, superconductors to excitonic insulators, and magnetic phases to ferroelectrics. This enormous range of behaviors together with a fairly universally applicable toolbox for assembly into heterostructures is providing nearly endless opportunities to create novel functionalities and new physics in the limit of extreme quantum confinement: Heterostructures have been demonstrated that support physics ranging from the anomalous quantum Hall effect[2] to unconventional superconductivity,[3] and from long-lived valley-polarized photoluminescence[4] to light-induced ferromagnetism,[5] to name just a few of the many surprising observations. Fundamentally, this is due to the increased importance of correlations in the 2D limit and their interplay with quantum fluctuations. As a result, there is a vigorous search underway for new van der Waals layered materials – we will refer to them as 2D materials even if outside the monolayer limit - to increase the available design space. Though this has been a very fertile area of research, fundamental limits remain: Only certain types of compounds and crystal structures lend themselves to van der Waals layering, and the effect of poor lattice matching in heterostructures may limit or suppress desired outcomes. In addition, many materials become chemically rather unstable in the 2D limit, making processing challenging.

In this Perspective, we pose the question whether the combination of van der Waals layered materials with organic molecules can increase the design space for creating well-defined hybrid organic/2D heterostructures that enable control of relevant electronic, spin and phononic degrees of freedom. In particular, we ask how such heterostructures enable tailoring *correlations* in the 2D limit, and discuss several mechanisms by which such control might proceed. This remains an important challenge in condensed matter physics since correlated phases such as charge-density waves, superconductivity and correlated insulator phases are often in competition with each other. The fundamental rationale for the proposed undertaking is the vast molecular library available from chemistry to create organic molecules with nearly infinitely tunable properties. Here we propose that hybrid organic/2D heterostructures offer indeed a largely unexplored path towards a new paradigm for studying correlated physics in low-dimensional structures. This is despite the fact that molecular adsorbates are often considered a nuisance when fabricating 2D heterostructures, introducing disorder and scattering centers. Our two take-away messages are as follows: 1) Organic adsorbates and intercalates can preserve, enhance and tailor correlation effects, making them a suitable alternative to all-inorganic heterostructures; and 2) in order to expand the potential of organic molecules beyond relatively simple modifications of the 2D material, it is useful to consider symmetries and to adopt the lens of proximity effects. We will explore each of these points in turn in what follows.

## II. Scope of the Perspective

The notion of combining 2D materials with organic molecules has a long history, particularly in the context of organic optoelectronics whose aim it is to design novel devices such as light-emitting diodes, photovoltaics, transistors etc. with organic molecules as active layers. Here, it has long been recognized that molecular arrangement and intermolecular interactions play a crucial role in device efficiency. Equally important are the electronic interactions at the device contacts, where charge is injected, extracted or transported. In this context, 2D materials have served as usually weakly interacting substrates that enable the study of these effects under highly controlled conditions.[6–15] New opportunities have arisen when combining organic molecules with 2D semiconductors, e.g. monolayer transition metal dichalcogenides (TMDs),[16,17] since these provide desirable properties such as direct bandgaps, and therefore enable ultrathin and flexible optoelectronics in their own right. Moreover, graphene and TMDs also exhibit unusually high surface-enhanced Raman scattering cross-sections, explored in some detail by Dresselhaus and us, among others.[18,19] Though rich and fascinating in its own right, the topic of how to understand the physics at play in such heterostructures designed for traditional optoelectronics is beginning to be quite well understood, and will not be discussed in this Perspective.

Organic molecules have also been used to stabilize inorganic mono- and few-layer materials,[20–22] creating *e.g.* crystalline organic molecule/2D superlattices in which layers of organic molecules alternate with layers of the 2D material in a highly ordered fashion. These composite materials are bulk in nature but reveal the properties of the underlying monolayer van der Waals materials in a protective and bulk-like matrix. These studies show the way to improved synthesis and materials control of layered materials but have so far not yet fundamentally altered the properties of their parent layered atomic crystals.

Instead, we shall concentrate here specifically on the question of whether and how organic molecules allow tailoring correlated states in van der Waals layered materials, both in the bulk and the 2D limit. The requirements for such a target are considerably more stringent than for optoelectronics, where one is principally concerned with understanding and engineering band alignments to enable charge and energy flow as needed for the optoelectronic function (light emission, light-to-charge conversion, current switching etc.). In contrast, one may expect that molecules in hybrid organic/2D materials need to be highly ordered so as to avoid distributions of scattering centers detrimental for electronic or phononic correlations, and that hybridization at the heterointerfaces must be neither too weak (no impact) nor too strong (damage to characteristics of 2D material), thereby destroying the quantum functionality. Some precedent in this arena exists already since the 1970s from studies of atomic intercalations, particularly in the context of tailoring superconductivity or the electronic structure,[23–25] suggesting that the assembly of hybrid heterostructures *e.g.* by intercalation might be a fruitful endeavor. We will show in the next section that *the scope for molecular design and targeted effect is in fact surprisingly large*, and that the *conditions for manipulating correlations are not overly stringent*, particularly in terms of structure of the organic thin film. Note that improved control of the organic self-assembly may very well be beneficial for tailoring the correlated states of the 2D materials. And yet, even though in many instances little is known about the specifics of the structure of the organic/2D heterointerface, there is clear evidence that organic molecules permit tailoring and even creating new correlated phases in organic/2D heterostructures.

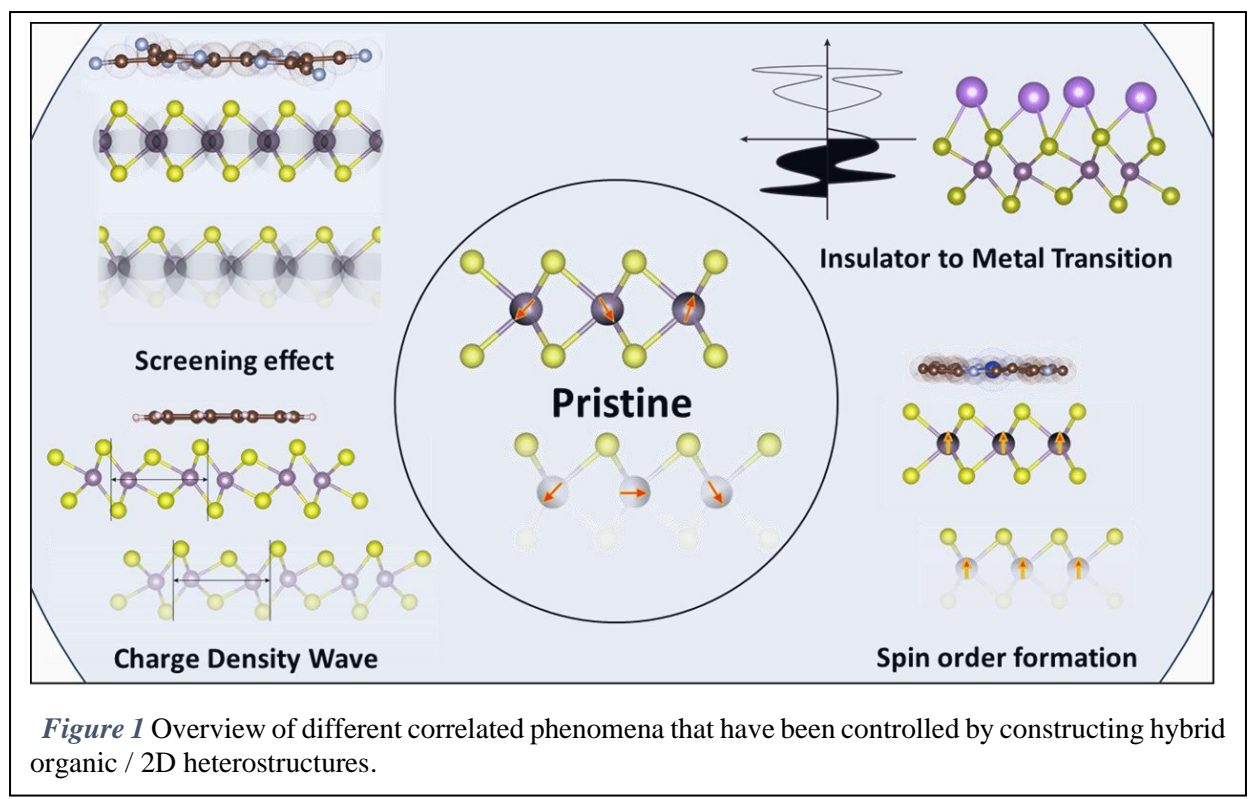

*Figure 1* Overview of different correlated phenomena that have been controlled by constructing hybrid organic / 2D heterostructures.

## III. Existing Reports

We will briefly review the limited number of reports of tailoring correlated phenomena by forming hybrid organic/2D heterostructures (**Figure 1**). We will group these reports by phenomenon in order to reveal common features and to highlight the observed magnitude of the effect. Though explanations offered for the different observations vary enormously and a unified picture is at present missing, this will allow us to bring to light unifying concepts and discuss the proposed underlying mechanisms in section IV. Readers primarily interested in a mechanistic understanding and a perspective on fruitful future directions and novices in the physics of 2D materials may wish to advance to section IV.

### A. Magnetism

Several studies have sought to manipulate the Curie temperature $T_C$ in layered ferromagnets. Huang *et al.* intercalated FeSe first with LiOH to create (Li,Fe)OHFeSe, a hybrid material with a superconducting phase below approximately 40 K.[26] Remarkably, intercalation with a substoichiometric fraction of the electron donor bis(ethylenedithio)tetrathiafulvalene (BEDT-TTF) results in a fundamentally different magnetic susceptibility, interpreted as a transition from the LiOH-intercalated paramagnetic state to a spin-glass-like state. Wang and coworkers electrochemically intercalated the ferromagnetic semiconductor $Cr_2Ge_2Te_6$ with tetrabutylammonium, inducing a transition to a metallic ferromagnet with a significantly increased Curie temperature from ~70 K to $T_C$ = 208 K.[27] Hueso et al. employed a thin film of Co-phthalocyanine (CoPc), grown in the anti-ferromagnetically coupled α-phase, on the metallic ferromagnet Fe$_3$GeTe$_2$ to take advantage of exchange bias that pins magnetization direction reversal to temperatures around 110 K.[28] Several density functional theory (DFT) studies suggest that doping of van der Waals magnets such as CrI$_3$ or MnPS$_3$ enhances the magnetocrystalline anisotropy. This opens a spin-wave excitation gap, thereby stabilizing magnetic order in low dimensions against the thermal fluctuations that by the Mermin-Wagner theorem are predicted to prohibit magnetic order.[29,30]

### B. Charge-Density Waves (CDW)

Kaiser and coworkers reported on the transition from the octahedral 1T polymorph of TaS$_2$ in the non-commensurate CDW phase to a mixed 2H/1T phase by intercalation of triethylenediamine.[31] They proposed that this transition is primarily driven by forcing a better overlap between the Ta and S orbitals, which promotes covalency and favors thus the 2H polymorph. Similar observations were reported earlier *e.g.* by intercalation with hydrazine, observing a transition from the non-commensurate to the commensurate CDW phase in TaS$_2$.[32,33] We note in passing that the inverse process, i.e. controlling molecules by CDW, has been reported as well: Self-assembly of manganese phthalocyanine (MnPc) is controllably altered by switching from the CDW 1T polymorph in NbSe$_2$ to the 2H polymorph.[34]

### C. Superconductivity

The most extensively investigated target for hybrid organic/2D materials heterostructures is the critical temperature $T_c$ for the superconducting phase transition. Early work by Geballe and subsequent efforts by others suggest that intercalation with small amines, ammonia, and some salts typically raises $T_c$.[35–38] Intercalation of MoTe$_2$ and WTe$_2$ with organic ionic liquids has been suggested to create a bulk material that consists of weakly coupled monolayers, consequently raising $T_c$.[39] A number of more recent studies suggest that $T_c$ may also be tailored by molecular adsorbates such as short-chain self-assembled monolayers and Pcs,[40,41] though the mechanism whereby this occurs is at present unclear.[42]

### D. Electronic Phase Transitions

A few studies suggest that organic layers may drive electronic phase transitions; *e.g.* sandwiching bilayer graphene between $NH_2$-terminated $SiO_2$ and a layer of 2,3,5,6-tetrafluoro-tetracyanoquinodimethane ($F_4TCNQ$) opens a bandgap.[43] Inspired by the large body of work on 2D heterostructures supporting moiré patterns, a one-dimensional potential modulation was created by assembling supramolecular lattices on graphene, and this potential could be photoswitched in a photoinduced molecular isomerization process.[44] Hybrid heterostructures also open up a possibility for new electronic phases, *e.g.* by creating a Kagome lattice by assembling a metal-organic framework on $NbSe_2$.[45]

### E. Other Types of Correlations

There are a number of reports that demonstrate manipulation of the spin degrees of freedom in organic/2D hybrids. Concerning spin *texture*, i.e. non-uniform arrangement of the spin direction in momentum space, Jakobs *et al.* showed that adsorption of a carefully designed phthalocyanine on the surface of the topological insulator $Bi_2Se_3$ creates distinct Rashba-split surface states.[46] A tantalizing suggestion by Zhu and colleagues indicates that disordered adsorption of hydrazine exerts strain[47,48] on superconducting $NbSe_2$, leading to a slight structural distortion of $NbSe_2$. This in turn reveals the magnetic moment of Nb, which may result in the coexistence of ferromagnetism and superconductivity.[49] This is a fascinating proposal since conventionally these two phases are considered to be mutually exclusive.

Liljeroth and coworkers assemble a 2D metal-organic framework (MOF) on $NbSe_2$, where the MOF forms a Kagome lattice.[45] The adsorption of CoPc dimers on superconducting $NbSe_2$ creates localized in-gap states (Yu-Shiba-Rusinov or YSR states) due to the magnetic moment of the molecules.[50] This suggests the possibility of assembling a magnetic MOF on $NbSe_2$ to create a YSR lattice, one proposed approach to achieve topological superconductivity. Remarkably, YSR states have also been suggested to appear when adsorbing helical peptides on $NbSe_2$,[51] in a manifestation of the chirality-induced spin selectivity effect (CISS; see section V.C for a brief introduction to CISS).[52]

## IV.   Proposed Mechanisms

The previous section convincingly demonstrates that organic adsorbates or intercalates may tailor a wide range of correlations in van der Waals materials, even when the structure of the heterointerface is not optimized, ordered or even understood. In some instances, such as magnetic phases, the effect can be significant, increasing critical temperatures by many tens to hundreds of Kelvin. For others, such as superconductors, the effect appears to be at present more modest. For yet others, fundamentally new phenomena such as phase transitions can be driven by forming an organic/2D heterostructure. This underlines our first message in this Perspective: *Organic adsorbates and intercalates harbor considerable promise for tailoring a wide range of correlated states in van der Waals materials.* What then is responsible for this potential?

*Table 1* Table of sample organic molecules for functionalizing 2D materials.

| FUNCTIONALITY | EXAMPLES |
| --- | --- |
| Electron withdrawing | HATCN, $F_4$TCNQ, PTCDA |
| Electron donating | TTF, CuPc |
| Spin functionality | $TbPc_2$, CoPc, MnPc, radicals |
| High dielectric constant | Ionic liquids |
| Intercalation | Amines, ammonium salts |

HATCN: Hexaazatriphenylenehexacarbonitrile; $F_4$TCNQ: 2,3,5,6-Tetrafluoro-tetracyanoquinodimethane; PTCDA: Perylenetetracarboxylic dianhydride; TTF: Tetrathiafulvalene; Pc: Phthalocyanine

Not surprisingly, many different explanations have been offered to help understand the broad range of observations of modification of correlated phenomena in layered and 2D materials (see **Table 1** for a list of common functionalities and sample molecules). Since electronic correlations involve excitations of quasiparticles near the Fermi surface, the most commonly proposed mechanism involves changes to the density of states near $E_F$ and/or doping of the 2D material. *E.g.* electrochemical intercalation of the semiconducting $Cr_2Ge_2Te_6$ results in strong doping and an insulator-to-metal phase transition, ultimately proposed to be responsible for the elevated Curie temperature.[27] Similar arguments were made for CoPc on $Fe_3GeTe_2$, consistent with gate-tunable ferromagnetism in this material.[53] In some instances, the specific doping mechanism was articulated very explicitly, typically derived from computational models: For doping of $CrI_3$ with TTF or $F_4TCNQ$ it was suggested that the Cr *d*-orbitals are depleted or filled, modifying the intersite exchange coupling accordingly.[29]

A very different picture of rehybridization is also occasionally proposed: Strain and structural distortions caused by adsorbates or intercalates may increase or reduce the degree of covalency between the transition metal and the chalcogenide in TMDs, thereby directly altering intersite coupling and thus the electronic correlations. This has been used to explain changes to the CDW phase in $TaSe_2$ and the proposed coexistence of ferromagnetism and superconductivity in $NbSe_2$ with a layer of hydrazine.[31,49]

In some instances, particularly involving intercalates with high dielectric constants such as ionic liquids, an increase in the $T_c$ for superconductivity may stem from layer-decoupling. In materials such as $MoTe_2$ and $WTe_2$, intercalation removes the 1T'→1T phase transition, effectively creating a monolayer-like material.[39] This is consistent with other reports of intercalation-driven layer decoupling,[24,54] and the fact that TMD monolayers typically exhibit higher phase transition temperatures.

Other, perhaps more subtle mechanisms can be imagined: Interfacial hybridization between the organic and the 2D material may introduce additional phonon modes or modify the existing phonon spectrum. One might predict that this would change the phase diagram of CDW phases and the critical temperature for superconductivity. For intercalated systems or organic/monolayer 2D heterostructures, the organic may also alter the dielectric environment. Consequently, changes in screening may enhance or suppress long-range interactions and the energy range of the coherence necessary for correlated phases to appear. For some phenomena, the balance of electron and hole pockets may be altered in the heterostructure, once again modifying the degree to which correlations can be supported. More trivially, scattering centers lead to the destruction of correlations, as has been proposed e.g. in the case of adsorbing magnetic MnPc on a superconducting monolayer of In.[41]

## V.    Novel Phenomena and New Opportunities

The brief summary of already explored phenomena and proposed mechanisms makes it clear that organic/2D heterostructures provide a rich test bed for correlated physics in reduced dimensions. Important questions remain: To what extent do organics need to form ordered structures for optimal manipulation of correlations? Are there specific relationships between the organic superlattice and the 2D material lattice that are advantageous for maximizing the ability to tailor correlations? What are the design principles for organics, beyond the broad strokes outlined in **Table 1**? Are there functional groups that are particularly advantageous for modifying certain types of 2D materials? The diversity of already reported observations undoubtedly deserves careful and detailed investigations to confirm, revise and deepen our understanding of critical energy scales and interaction mechanisms in such heterostructures, and future research should address these questions.

In the remainder of this Perspective we will instead aim to establish a shifted view point that we suggest is useful to explore new phenomena that may be uniquely or ideally accessible with organic/2D heterostructures. We will start by first briefly highlighting the extent to which fabrication of such heterostructures appears to be quite unproblematic and considerably less complex than more conventional 2D/2D heterostructures. This practical advantage holds promise for rapid advances in creating the diverse array of heterostructures needed to generate a deeper understanding of the interactions at the organic/2D interface and their energy scales.

We will then turn our attention to novel possibilities to tailor correlated phenomena in 2D and in layered materials that are at present only poorly explored by introducing the concept of "proximitization". We will show that this atomic-scale view of coupling of molecular degrees of freedom to the 2D material together with symmetry considerations provides new opportunities to manipulate correlation effects in van der Waals materials.

### A. The organic advantage

Fabrication of 2D/2D heterostructures is often complicated and cumbersome: Individual flakes must be found, transferred and stacked, sometimes with extraordinary precision to achieve the desired effect. This is different for organic/2D heterostructures: Though one loses the twist angle between 2D sheets as a valuable degree of freedom,[55] organic/2D material heterostructures gain from the fact that lattice registry between the organic and the 2D lattices (epitaxy) does not appear to be needed to considerably influence correlations in the 2D material, as is apparent from some of the studies highlighted in Section III. Indeed, these heterostructures can be created in a straightforward manner, benefitting from the considerable body of knowledge in organic molecular beam epitaxy.[56] At this point in time, the main consideration appears to be the choice of molecule and molecular functionalization, with one fundamental limitation already evident: Molecules that do react vigorously by making and breaking bonds with the 2D material typically destroy the van der Waals material, and as a consequence the ability to manipulate correlations. One might expect that on the opposite end of the interaction scale, molecules that interact only weakly via van der Waals interactions leave the 2D material largely unaffected. Remarkably though, this is not the case: Not only can well-defined organic/2D materials heterostructures be formed, but new functionalities can be imparted. Beyond some of the examples discussed in Section III, we will highlight an example in more detail in Section V.D. Consequently, restrictions on fabrications are mild, and many different combinations can therefore be envisaged and likely grown with ease. In many instances, the many organic adsorbates also act as passivation layers vis-à-vis ambient reactants such as oxygen and water, enhancing the stability of the 2D material. Most importantly, the vast design space of organic molecules allows one to deliberately and rationally tailor the properties of the molecular adsorbate, and hence potentially of the complete heterostructure.

B. Proximitization

By "proximitization" we mean the imposition of a physical phenomenon in one material by the atomic length-scale contact with a second, usually different material (**Figure 2**).[57] This may simply transfer properties from one material to another in the interfacial region, or give rise to entirely new phenomena. A well-established example of proximitization is the penetration of superconductivity from a superconductor into a normal metal, e.g. in the case of amorphous W into Co.[58] In this case, the mechanism is well-understood and involves the formation of Cooper pairs upon Andreev reflection at the superconductor-normal interface. Generally, proximity effects are rather short range, commonly on the order of 1-100 Å. The fact that layers in van der Waals layered materials are atomically thin with thicknesses between 3 and 10 Å per layer implies that these materials are thus ideally suited for harnessing proximitization: Interactions imposed or generated by the organic adsorbate layer are restricted to one to a few layers in the van der Waals material, either transforming the few-layer 2D material or creating a quasi-2D material in the interfacial region of the heterostructure with a bulk van der Waals material. Only the first few layers will experience the proximity field, and the interface established in this way creates a novel type of proximitized bulk material whose near-interfacial region – the first few layers - has profoundly different properties. This is radically different from most other 3D bulk materials where coupling between different lattice planes is much stronger than in van der Waals materials. As a result, even short-range proximity effects can be expected to significantly modify the properties of van der Waals materials. A recent manifestation of this idea can be found in the enhancement of spin-orbit coupling (SOC) in graphene by three orders of magnitude through proximitization with $WSe_2$ or with a topological insulator such as $Bi_2Se_3$,[59,60] both of which have strongly spin-orbit split bands.

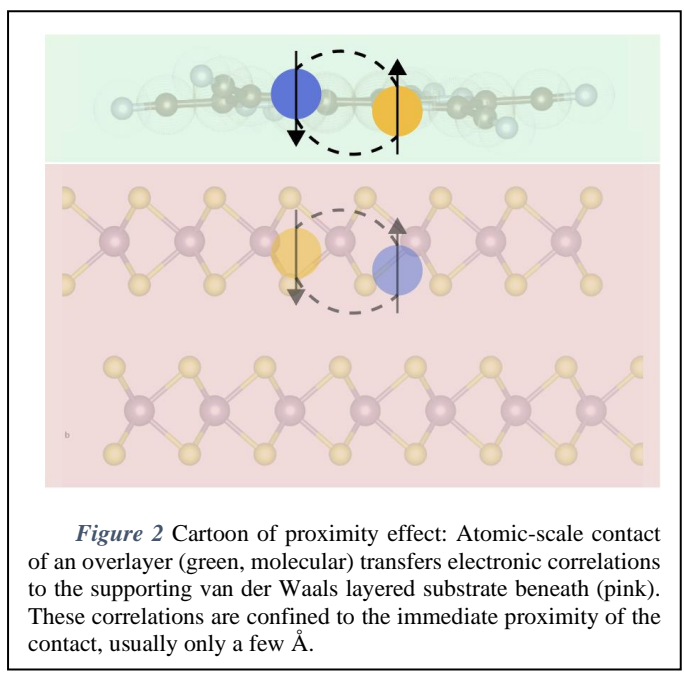

*Figure 2* Cartoon of proximity effect: Atomic-scale contact of an overlayer (green, molecular) transfers electronic correlations to the supporting van der Waals layered substrate beneath (pink). These correlations are confined to the immediate proximity of the contact, usually only a few Å.

"Conventional" proximitization of van der Waals materials, i.e. in combination with inorganic materials, is beginning to receive some attention, in particular for the case of graphene. *E.g.*, a recent proposal suggests that a van der Waals interface of nonmagnetic graphene with ferromagnetic Co, separated by a layer of hBN, results in graphene bands that are spin-split by more than 10 meV; i.e., the magnetic field of Co penetrates as a proximity field into graphene and spin-polarizes its bands.[61,62] Remarkably, gating is suggested to invert the spin polarization. Such effects are not possible with conventional ferromagnetic metals, demonstrating the unique properties of proximitized materials systems. Perhaps even more strikingly, proximity-induced SOC in graphene proximitized with transition metal dichalcogenides can lead to non-trivial band inversions and hence the emergence of topologically non-trivial states of matter,[63] with suggestions of the emergence of helical edge states.[64,65]

### C. Opportunities for proximitization in organic/2D heterostructures

Where then lie the opportunities for using proximitization and manipulation of correlated states with organic/2D material heterostructures (**Figure 3**)? As a starting point, some functionalities are unique to molecules and may thus offer some of the clearest views of the potential of proximitized

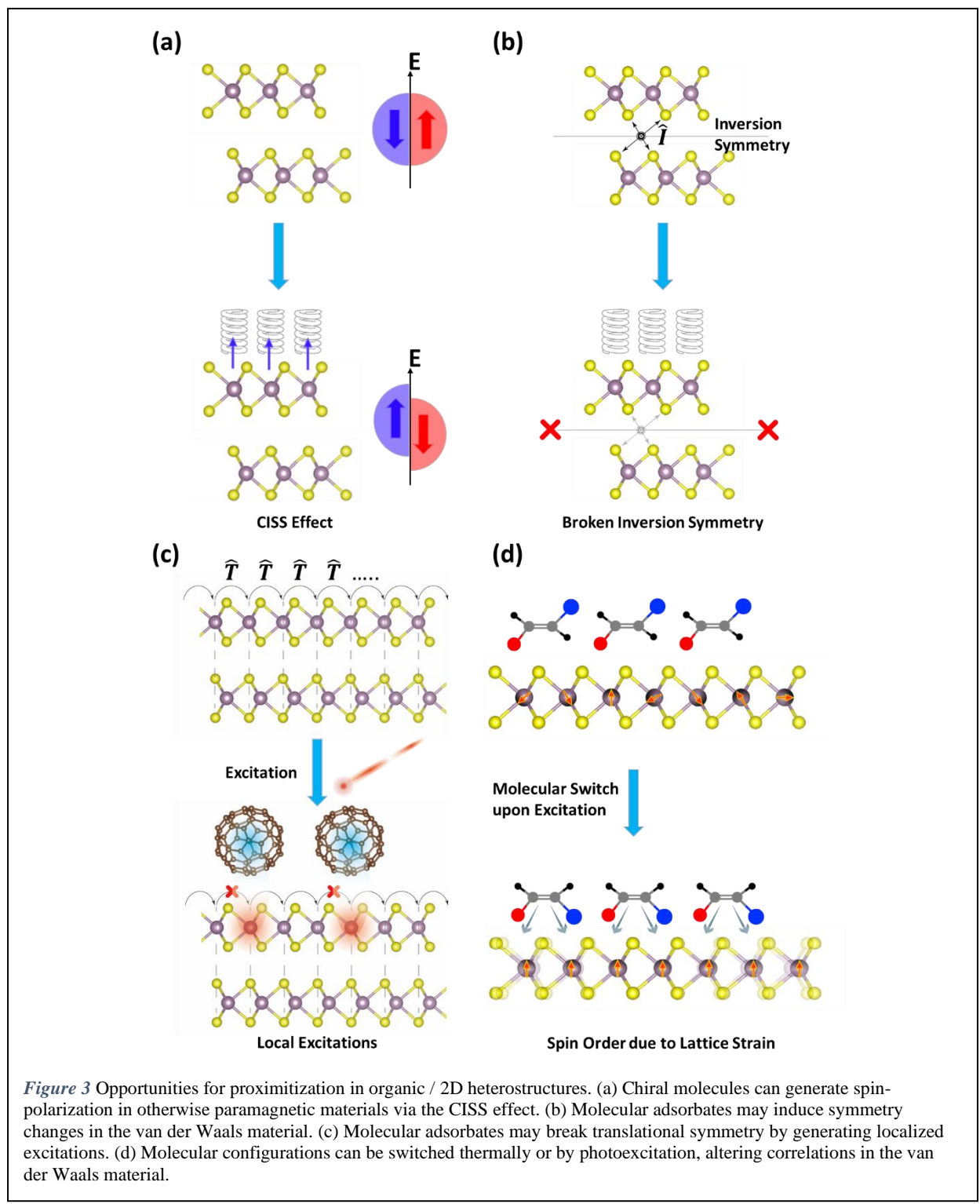

*Figure 3* Opportunities for proximitization in organic / 2D heterostructures. (a) Chiral molecules can generate spin-polarization in otherwise paramagnetic materials via the CISS effect. (b) Molecular adsorbates may induce symmetry changes in the van der Waals material. (c) Molecular adsorbates may break translational symmetry by generating localized excitations. (d) Molecular configurations can be switched thermally or by photoexcitation, altering correlations in the van der Waals material.

organic/2D heterostructures. E.g., certain molecules undergo isomerization under thermal or optical stimulus. This means that the molecular structure or molecular properties can be changed reversibly and on demand, *e.g.* by breaking or forming bonds such as in photochromism of rhodamine dyes,[66] by photoinduced rotation about a double bond in a cis-trans isomerization *e.g.* in azobenzene derivatives,[67] or in spin cross-over compounds where the spin state of a molecular transition metal complex is switched by an external stimulus.[68] The potential to collectively switch an array of molecules in immediate proximity to a 2D material may make several desirable effects available: 1) The concerted switching of the molecular adsorbate structure may exert a strain on the 2D material. Lattice strains of just a few % are known to have a profound impact on the electronic properties of the 2D material, *e.g.* opening and closing gaps in electronic bands[69,70] or inducing ferromagnetism[47] by altering bonding interactions in the lattice of the 2D material. 2) Manipulating the ligand field of a transition metal in a transition metal complex by an external stimulus can drive the spin configuration of the *d*-electrons in the transition metal e.g. from an all spins paired low-spin configuration with zero magnetic moment to a high-spin state with a significant magnetic moment. By proximity to the 2D material, this may establish a magnetic proximity field which in turn may lift the spin degeneracy in the 2D material. Particularly promising for all these and related effects is the fact that photo-driven isomerization processes occur on ultrafast timescales. This may therefore make it possible to drive control of correlated phases with unprecedented ultrafast speeds.

A second important unique advantage of molecules is the fact that many molecules are chiral. Chirality in molecules means that a molecule cannot be superimposed with its mirror image; hence, the molecule lacks inversion symmetry. Breaking of symmetries is necessary to support certain correlated phases, most obviously in the case of broken time-reversal symmetry in a ferromagnet. Proximitization with chiral molecules may therefore hold the key for enabling and manipulating a range of phenomena that rely on broken inversion symmetry. *E.g.*, in the presence of strong SOC, broken inversion symmetry creates spin and even orbital texture in the Bychkov-Rashba effect.[71]

This means that the spin or orbital angular momentum direction depends on $\vec{k}$, the vectorial electron momentum. It is thus conceivable to employ chiral molecules to develop, amplify or manipulate inversion symmetry breaking in proximity to a 2D material with SOC. Manipulating spin texture offers an avenue to the manipulation of spin for spintronics applications, e.g. by spin-to-charge conversion or by spin-orbit torque. Also reliant on broken inversion symmetry is the phenomenon of ferroelectricity, already observed in certain low-dimensional van der Waals materials, and promising for new approaches to fabricating non-volatile memory.[72] It may be possible to use chiral molecules to manipulate the Curie temperature for ferroelectric order in a proximitized organic/2D heterostructure. As a final example, in addition to band inversion, inversion symmetry breaking is also a key ingredient for the emergence of topologically non-trivial Weyl semimetal phases.[73] Weyl semimetals host Weyl nodes, points in momentum space where electrons behave as robust relativistic Fermions with magnetic monopole-like signatures. Though at present most reported Weyl semimetals are 3D bulk materials, proposals for 2D Weyl heterostructures exist.[74] These have not yet been experimentally realized, and it may be possible to advance this search for 2D Weyl semimetals by breaking inversion symmetry via proximitization in an organic/2D heterostructure.

Molecular chirality is also drawing significant attention due to chirality-induced spin-selectivity (CISS).[52] CISS manifests as the selective transport of spin up over spin down electrons through helical or chiral molecules in the absence of an external magnetic field or a magnetic substrate. This is a remarkable observation, and even though the mechanism by which CISS acts remains to be fully understood, the evidence that chiral/helical molecules give rise to spin-polarized transport is solid.[75,76] Some recent reports suggest that chiral molecules additionally support a proximity magnetic field that can be used *e.g.* for magnetic imaging and magnetization switching.[77,78] This opens the door for organic/2D heterostructure designs that take advantage of this uniquely molecular effect, *e.g.* with 2D materials that reside close to the Stoner instability where a proximity field might have a large influence, or in creating time-reversal symmetry broken phases of quantum matter.

The unifying theme in proximitization with organic molecules is the concept of symmetry breaking: Organic molecules confer their properties at the heterostructure interface in such a way as to break or enhance the breaking of already broken symmetries in the van der Waals material. Beyond inversion and time-reversal symmetry breaking, molecular adsorbates may break translational symmetry even when forming ordered superlattices. This notion stems from the fact that molecular thin films and solids usually form insulating phases with very weak intersite hopping even when highly ordered. As a result, molecular excitations (magnetic, electronic, excitonic) remain over extended duration strongly localized to one or at most a few lattice sites, thereby temporarily breaking translational symmetry. This is in contrast to conventional inorganic / inorganic heterostructures, many of which host strongly dispersing bands and therefore delocalized electronic states. As a consequence, excitations carry the signatures of the translationally symmetric wavefunctions, delocalized over the full lattice over all but the shortest times. This distinction is important, since the proximity fields emanating from localized excitations in organic/2D heterostructures may be remarkably strong, creating *e.g.*, large local electric or magnetic fields at the heterostructure interface. We highlight this idea with a case study from our laboratory in the next section.[79] Overall, this constitutes our second main message of this Perspective: *Organic adsorbates offer a rich toolbox with which symmetries in the 2D materials can be broken by proximity-induced effects*.

D. A case study: Time-reversal symmetry breaking by localized proximity fields

The combination of an organic molecule with a TMD constitutes the assembly of a semiconductor heterojunction. Upon excitation, and depending on the relative band alignment of the respective valence and conduction bands, it is possible to drive interfacial charge-transfer, *e.g.* from the organic to the TMD. As discussed above, due to the weak intermolecular interactions, the resulting hole on the organic is expected to remain localized at short times to one or a small number of molecules, transiently breaking translational symmetry if the molecular adsorbate is ordered. This creates a large localized electric field across the interface, which, as we will discuss, breaks inversion symmetry in the TMD. One may then take advantage of the unique properties of TMDs (**Figure 4**): In semiconducting TMDs and due to large SOC, the valence band near the corner of the hexagonal Brillouin zone (*K* point or *valley*) exhibits large spin-splitting, linking spin and valley. Of course, across the complete unit cell the net spin polarization is zero, as expected for a non-magnetic material. In TMDs of the 2H polytype, the order of the spin-split valence bands inverts between top and bottom layer in the unit cell due to the inversion symmetry of the bilayer unit cell. This couples the layer to the spin and valley as well. We thus speak of spin-valley-layer locking near ±*K* in 2H-TMDs. Consequently, translational symmetry breaking in the heterostructure also breaks the inversion symmetry, which in turn not only lifts the layer degeneracy but also the spin degeneracy of the valence bands, thus creating an opportunity to *create* spin polarization without external magnetic fields.

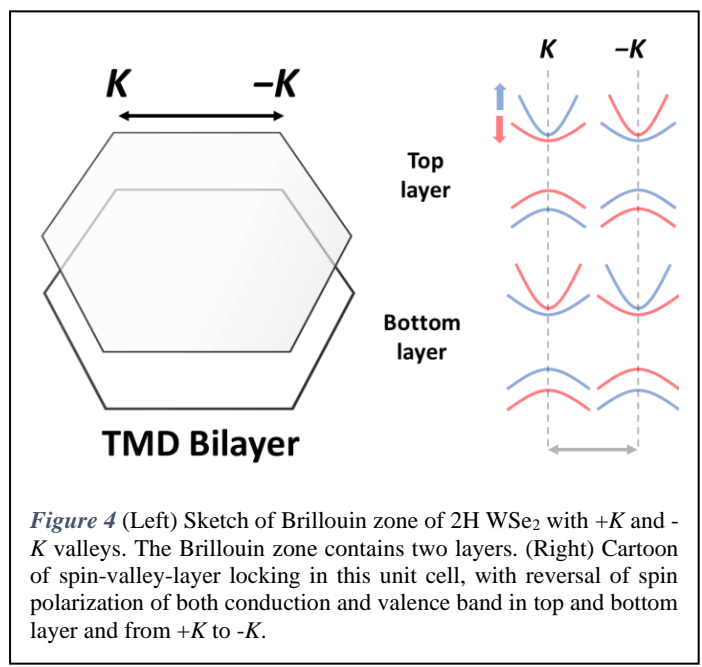

*Figure 4* (Left) Sketch of Brillouin zone of 2H WSe$_2$ with +*K* and -*K* valleys. The Brillouin zone contains two layers. (Right) Cartoon of spin-valley-layer locking in this unit cell, with reversal of spin polarization of both conduction and valence band in top and bottom layer and from +*K* to -*K*.

We chose to investigate this scenario for the case of $C_{60}$ on 2H-WSe$_2$, since compared to SOC, interlayer hopping in WSe$_2$ is small, substantially decoupling individual layers in WSe$_2$.[80] We followed the evolution of the electronic structure and creation of spin polarization in this heterostructure by time-, spin- and angle-resolved photoemission. Upon excitation of $C_{60}$ with an ultrafast UV pulse, a short-lived molecular excitonic state undergoes interfacial charge-transfer to the *K* point of the conduction band of WSe$_2$. The electron scatters away from *K* into the $\Sigma$ valley of the WSe$_2$ conduction band within approximately 1 ps, and remains there for almost 10 ps. Meanwhile, the hole remains largely localized on a single $C_{60}$ molecule. The interfacial charge-transfer and localized photohole in $C_{60}$ establish therefore a strong and localized directional electric field (~$10^7$ V/cm) normal to the $C_{60}$/WSe$_2$ interface, effectively breaking the $C_{60}$ translational symmetry, and, more importantly, the inversion symmetry of the bilayer unit cell of 2H-WSe$_2$. This is because the strength of this interfacial field decays into the WSe$_2$ crystal and is strongest in the top layer. Indeed, while the electronic wavefunction in the *K* valley is localized to a single WSe$_2$ sheet in the bilayer unit cell of 2H-WSe$_2$, it is delocalized over both layers in the $\Sigma$ valley. Consequently, one expects an electronic structure modification of the top layer of WSe$_2$ over approximately 1 ps, weakened thereafter as the electron begins to delocalize into both layers, and decaying after approximately 10 ps. This is indeed the case: Both the $C_{60}$ and the WSe$_2$ bands shift by approximately 50 meV over this time window. This shift can be understood as a dynamic Stark shift, induced by the transient proximity field established by the localized interfacial charge-transfer.

The spin-valley-layer locking of 2H-WSe$_2$ near the $K$-points can be described by a simple Hamiltonian, $H = -\lambda \tau_z s_z \sigma_z + t_\perp \sigma_x$, which captures the SOC-mediated ($\lambda$) interaction of the spin degree of freedom $s_z$ with the valley and layer pseudo-spins $\tau_z$ and $\sigma_z$ in the presence of weak interlayer hopping $t_\perp$. This spin-valley-layer locking implies that manipulation and polarization of one of the three degrees of freedom (spin, valley pseudo-spin or layer pseudo-spin) results in polarization of the others. In the case of charge-transfer from C$_{60}$ to the $K$-valley of WSe$_2$, the electric field created by the charge-separation is layer-dependent due to the initial electron localization in the top layer. Such layer dependence constitutes polarization of the layer pseudo-spin, which ultimately translates to layer-dependent spin polarization and lifts the spin degeneracy between the first and second layer in the top two layers of the WSe$_2$ crystal. The net result is the establishment of a ferromagnetic-like layer-dependent spin polarization. At its core, the electric field created by the interfacial charge-transfer breaks the inversion symmetry in 2H-WSe$_2$ and reveals the spin polarization hidden in the bilayer unit cell.[81]

From the context of organic / 2D heterostructures, these results are remarkable for two reasons: First, optically stimulated interfacial charge-transfer enables the manipulation of spin degrees of freedom on fs timescales, an important milestone for spintronics, realized here by taking advantage of the unique features of such organic/2D heterostructures. Second, unlike most ultrafast magnetization dynamics studies, rather than *destroying* magnetization and studying its recovery, here we are able to *create* spin polarization. Creation of ultrafast spin polarization constitutes a central challenge in spintronics, and it is proximitization in the unconventional nature of the organic/2D material heterostructure that makes this possible for the first time.

This case serves as a specific instance of the broader view offered by this Perspective: Organic/2D heterostructures can be readily grown, and organic molecules allow imparting new properties to the 2D or van der Waals material. In addition to explicitly modifying electronic and phononic densities of state, the organic film may establish and support a short-range field that breaks symmetries in the van der Waals material, thereby manipulating or revealing new correlations.

## VI. Concluding remarks

Though by many measures still in its infancy, there is an increasing body of evidence demonstrating that the combination of organic semiconductors with van der Waals layered materials offers unique opportunities and advantages for controlling and tailoring correlated phenomena in reduced dimensionality. Possibilities range from instances where molecules "merely" offer a wider design space for what can be already achieved in purely inorganic heterostructures, to novel phenomena that can only be obtained in hybrid heterostructures. Despite the fact that mechanisms of interaction remain to be explored in more detail, it is clear that electronic interactions play an important role. In this Perspective we propose broadening this viewpoint by considering design principles from the vantage point of proximity effects. Atomically thin layered materials are ideally suited to this purpose, and molecular adsorbates offer rich opportunities for proximitization.

## Acknowledgments

Discussions with Igor Žutić (University at Buffalo) are gratefully acknowledged. This work was supported by the Air Force Office of Scientific Research under award # FA9550-21-1-0219. JP, AB and OLAM also gratefully acknowledge support from the U.S. National Science Foundation award # CHE-1954571.

[24] Eads, C. N., Zachritz, S. L., Park, J., Chakraborty, A., Nordlund, D. L., and Monti, O. L. A. "Ultrafast Carrier Dynamics in Two-Dimensional Electron Gas-like K-Doped MoS2" *J. Phys. Chem. C* 124, no. 35 (2020): 19187–19195. doi:10.1021/acs.jpcc.0c07079

[25] Kang, M., Kim, B., Ryu, S. H., Jung, S. W., Kim, J., Moreschini, L., Jozwiak, C., Rotenberg, E., Bostwick, A., and Kim, K. S. "Universal Mechanism of Band-Gap Engineering in Transition-Metal Dichalcogenides" *Nano Lett.* 17, no. 3 (2017): 1610–1615. doi:10.1021/acs.nanolett.6b04775

[26] Huang, Y., Wolowiec, C., Zhu, T., Hu, Y., An, L., Li, Z., Grossman, J. C., Schuller, I. K., and Ren, S. "Emerging Magnetic Interactions in van Der Waals Heterostructures" *Nano Lett.* 20, no. 11 (2020): 7852–7859. doi:10.1021/acs.nanolett.0c02175

[27] Wang, N., Tang, H., Shi, M., Zhang, H., Zhuo, W., Liu, D., Meng, F., Ma, L., Ying, J., Zou, L., Sun, Z., and Chen, X. "Transition from Ferromagnetic Semiconductor to Ferromagnetic Metal with Enhanced Curie Temperature in Cr2Ge2Te6 via Organic Ion Intercalation" *J. Am. Chem. Soc.* 141, no. 43 (2019): 17166–17173. doi:10.1021/jacs.9b06929

[28] Jo, J., Calavalle, F., Martín-García, B., Tezze, D., Casanova, F., Chuvilin, A., Hueso, L. E., and Gobbi, M. "Exchange Bias in Molecule/Fe3GeTe2 van Der Waals Heterostructures via Spinterface Effects" *Advanced Materials* 34, no. 21 (2022): 2200474. doi:10.1002/adma.202200474

[29] Tang, C., Zhang, L., and Du, A. "Tunable Magnetic Anisotropy in 2D Magnets via Molecular Adsorption" *J. Mater. Chem. C* 8, no. 42 (2020): 14948–14953. doi:10.1039/D0TC04049E

[30] Wang, K., Ren, K., Cheng, Y., Chen, S., and Zhang, G. "The Impacts of Molecular Adsorption on Antiferromagnetic MnPS3 Monolayers: Enhanced Magnetic Anisotropy and Intralayer Dzyaloshinskii–Moriya Interaction" *Mater. Horiz.* 9, no. 9 (2022): 2384–2392. doi:10.1039/D2MH00462C

[31] Kinyanjui, M. K., Holzbock, J., Köster, J., Singer, C., Krottenmüller, M., Linden, M., Kuntscher, C. A., and Kaiser, U. "Spectral and Structural Signatures of Phase Transformation in the Charge Density Wave Material $1T\text{-}\mathrm{Ta}{\mathrm{S}}_{2}$ Intercalated with Triethylenediamine" *Phys. Rev. B* 103, no. 6 (2021): 064101. doi:10.1103/PhysRevB.103.064101

[32] Ghorayeb, A. M. and Friend, R. H. "Transport Evidence for New Phase Changes in 1T-TaS2 after Intercalation with Hydrazine" *J. Phys.: Condens. Matter* 6, no. 19 (1994): 3533. doi:10.1088/0953-8984/6/19/008

[33] Tatlock, G. J. and Acrivos, J. V. "In Situ Intercalation of TaS2 in the Electron Microscope" *Philosophical Magazine B* 38, no. 1 (1978): 81–93. doi:10.1080/13642817808245322

[34] Zhang, Q., Huang, Z., Hou, Y., Yuan, P., Xu, Z., Yang, H., Song, X., Chen, Y., Yang, H., Zhang, T., Liu, L., Gao, H.-J., and Wang, Y. "Tuning Molecular Superlattice by Charge-Density-Wave Patterns in Two-Dimensional Monolayer Crystals" *J. Phys. Chem. Lett.* 12, no. 14 (2021): 3545–3551. doi:10.1021/acs.jpclett.1c00230